\documentclass[preprint,number,sort&compress]{elsarticle}

\usepackage{amsmath,amssymb} 

\usepackage{hyperref}

\journal{Journal of \LaTeX\ Templates}

\begin{document}

\begin{frontmatter}

\title{Predictive modeling of passive scalar transfer to a wall using stochastic one-dimensional turbulence}

%
%

\author[addr1]{Marten Klein\corref{cor1}} 
\ead{marten.klein@b-tu.de}

\author[addr1]{Heiko Schmidt} 

\address[addr1]{Lehrstuhl Numerische Str\"omungs- und Gasdynamik, Brandenburgische Technische Universit\"at Cottbus-Senftenberg, Siemens-Halske-Ring 15A, 03046 Cottbus, Germany}

\cortext[cor1]{Corresponding author.
  Tel.: +49-(0)355-695-127;
  Fax: +49-(0)355-694-891.
}

\begin{abstract} 
Passive scalars in turbulent channel flows are investigated as canonical problem for heat and mass transfer in turbulent boundary-layer flows.
The one-dimensional turbulence model is used to numerically investigate the Schmidt and Reynolds number dependence of the scalar transfer to a wall due to fluctuating wall-normal transport.
First, the model is calibrated for low-order velocity statistics.
After that, we keep the model parameters fixed and investigate low-order passive scalar statistics for a relevant Schmidt and Reynolds number range.
We show that the model consistently predicts the boundary layer structure and the scaling regimes, for which it is close to asymptotic one-dimensional theory. 
\end{abstract}


\end{frontmatter}


\section{\label{sec:intro} Introduction}

Numerical modeling of scalar transport in turbulent boundary layers is a standing challenge that is relevant for applications from the technical to the atmospheric scales.
Key problems are related to small-scale correlations, scale interactions, counter-gradient fluxes, and numerical filtering (e.g.~\cite{Pirozzoli_etal:2016,Abe_Kawamura_Matsuo:2004,Hasegawa_Kasagi:2009}). 
Due to the latter, all relevant scales of the flow have to be resolved for robust numerical predictions.
Direct numerical simulation (DNS) would be the ideal tool, but it is of limited applicability due to the resolution requirements imposed by the Kolmogorov and Batchelor scales (e.g.~\cite{Hasegawa_Kasagi:2009,Schwertfirm_Manhart:2007}). 

\begin{figure}[t]
  \centering
  \includegraphics[height=38mm]{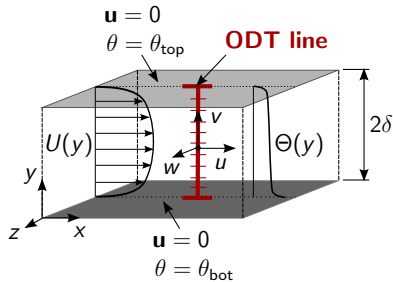} 
  \caption{%
    Schematic of the channel flow set-up investigated.
    ODT simulations are carried out for a lower-order computational domain, the so-called `ODT line'.
  }
  \label{fig:config}
\end{figure}

We address the numerical challenge of small-scale resolution by utilizing the stochastic one-dimensional turbulence (ODT) model \cite{Kerstein:1999,Kerstein_etal:2001}.
This model has been validated from a fundamental point of view and is now applied to multi-physics boundary layers (e.g.~\cite{Shihn_DesJardin:2007,Medina_etal:2019,Rakhi_etal:2019}). 
Here we apply the model to canonical channel flow as sketched in figure~\ref{fig:config} and investigate the Schmidt and Reynolds number dependence of the scalar transfer to a wall.
We limit our attention to the transport of passive scalars that have no effect on the mass, momentum, and energy balances of the flow. 

The rest of this paper is organized as follows.
In section~\ref{sec:method} we give an overview of the ODT model.
In section~\ref{sec:results} we report and discuss scaling regimes of the scalar transfer coefficient, the boundary-layer structure, and the turbulent eddy diffusivity.
At last, in section~\ref{sec:conc}, we close with our conclusions.

\section{\label{sec:method} Method}


\subsection{\label{sec:odt} Overview of the ODT model formulation}

Kerstein's \cite{Kerstein:1999,Kerstein_etal:2001} one-dimensional turbulence model aims to resolve all relevant scales of a turbulent flow. 
This is made feasible for high Schmidt and Reynolds numbers by modeling the effects of turbulent eddies by a stochastic process.
The model directly resolves the deterministic (molecular-diffusive) transport processes but models turbulent advection along a notional line-of-sight through the turbulent flow, the so-called `ODT line' as sketched in figure~\ref{fig:config}.
Hence, there is no closure and no closure modeling involved.

Here we consider constant-property channel flows with a passive scalar that has the same mass density as the bulk of fluid. 
The lower-order stochastic equations that describe such flows may be written as
\refstepcounter{equation}
$$
  \frac{\partial \boldsymbol{u}}{\partial t} + \mathcal{E}_{\boldsymbol{u}}(\boldsymbol{u}) = \nu \frac{\partial^2 \boldsymbol{u}}{\partial y^2} - \frac{1}{\rho} \frac{\mathrm{d}P}{\mathrm{d}x}\,\boldsymbol{e}_x \;,
  \qquad
  \frac{\partial \theta}{\partial t} + \mathcal{E}_\theta(\boldsymbol{u}) = \Gamma \frac{\partial^2 \theta}{\partial y^2} \;,
  \eqno{(\theequation{\mathit{a},\mathit{b}})}
  \label{eq:gov}
$$
where $\boldsymbol{u}=(u,v,w)^\mathrm{T}$ denotes the velocity vector and its Cartesian components, $\theta$ the scalar concentration, $t$ the time, $x$ the streamwise and $y$ the wall-normal (model-resolved) coordinate,  $\rho$ and $\nu$ the fluid's density and kinematic viscosity, $\Gamma$ the scalar diffusivity, $\mathrm{d}P/\mathrm{d}x$ the prescribed mean pressure gradient, $\boldsymbol{e}_x$ the unit vector in streamwise direction, and $\mathcal{E}_{\boldsymbol{u}}(\boldsymbol{u})$ and $\mathcal{E}_\theta(\boldsymbol{u})$ the effects of stochastic eddy events for the velocity vector and the scalar, respectively.
Note that $\mathcal{E}_{\boldsymbol{u}}$ and $\mathcal{E}_\theta$ are coupled and depend on the momentary velocity profile, $\boldsymbol{u}(y,t)$.
The largest permissible eddy size, $l$, is the channel half-height, $l\leq\delta$ \cite{Schmidt_etal:2003}.
At last, no-slip and Dirichlet wall-boundary conditions are prescribed for the velocity vector and the scalar, respectively.
See \cite{Kerstein:1999,Kerstein_etal:2001,Lignell_etal:2013,Klein_etal:2019} for further details. 

Similarity solutions to equations~(\ref{eq:gov}\textit{a,b}) are obtained in dependence on the Schmidt (Prandtl), $Sc=\nu/\Gamma$, and friction Reynolds, $Re_\tau=\delta u_\tau/\nu$,  number, where $\delta$ is the channel half-height, $u_\tau=\left(\nu\,\left|{\mathrm{d}U}/{\mathrm{d}y}\right|_\mathrm{w}\right)^{1/2}$ the friction velocity, $U=\bar{u}$ the mean velocity, and the subscript `$\mathrm{w}$' indicates evaluation at the wall.

\subsection{\label{sec:val} Remarks on the ODT model application and calibration}

ODT simulations of turbulent channel flows are conducted as follows. 
Equations~(\ref{eq:gov}\textit{a,b}) are numerically integrated to yields a time sequence of synthetic but statistically representative flow profiles on an adaptive grid \cite{Lignell_etal:2013}.
Conventional statistics are gathered for these profiles on a predefined post-processing grid.
The computation of cumulative statistics is straightforward, but the ODT-resolved turbulent fluxes are obtained by conditional eddy-event statistics \cite{Klein_etal:2019}.

\begin{figure}[t]
  \centering
  \includegraphics[height=42mm]{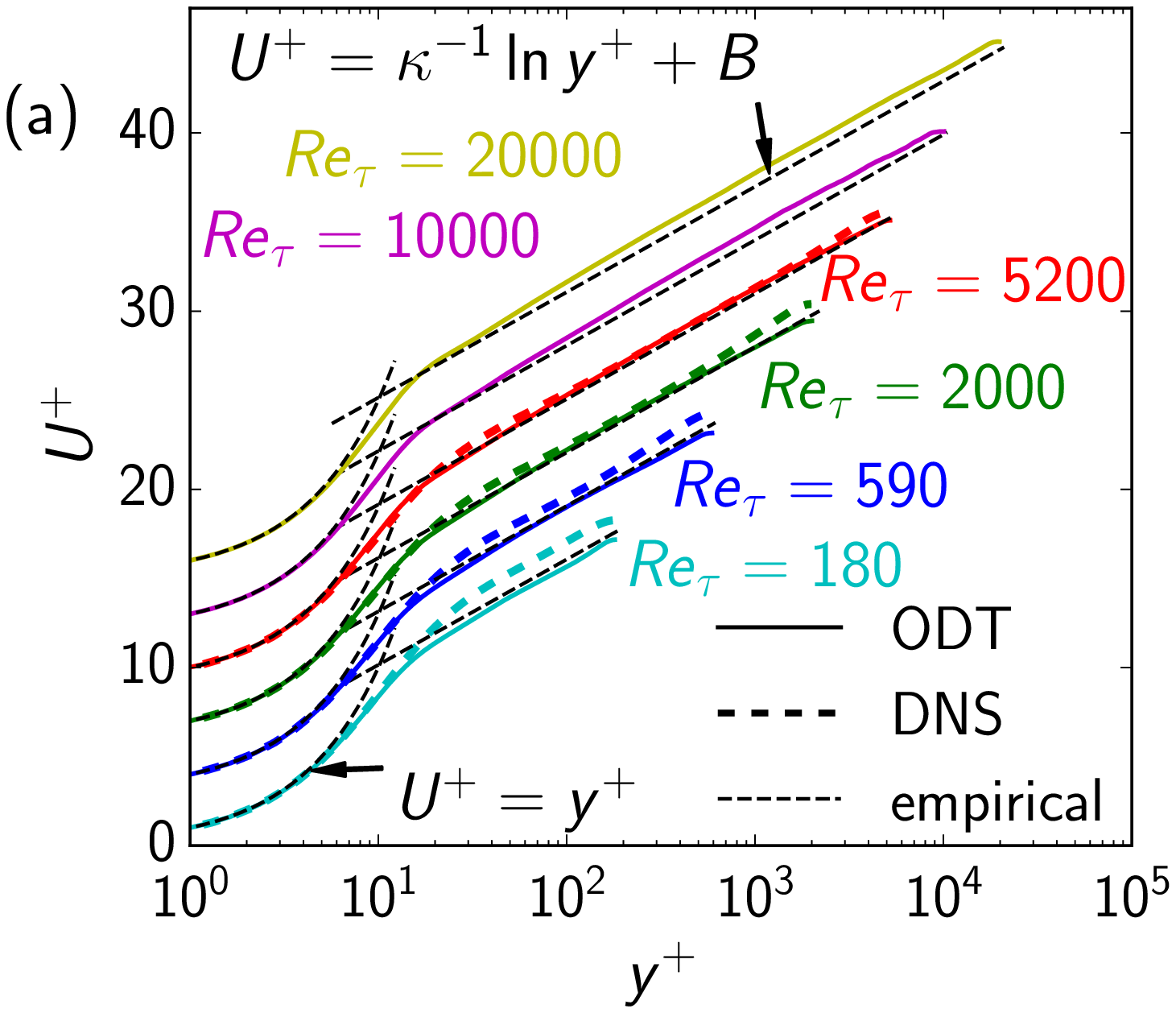} \hspace*{-3mm}
  \includegraphics[height=42mm]{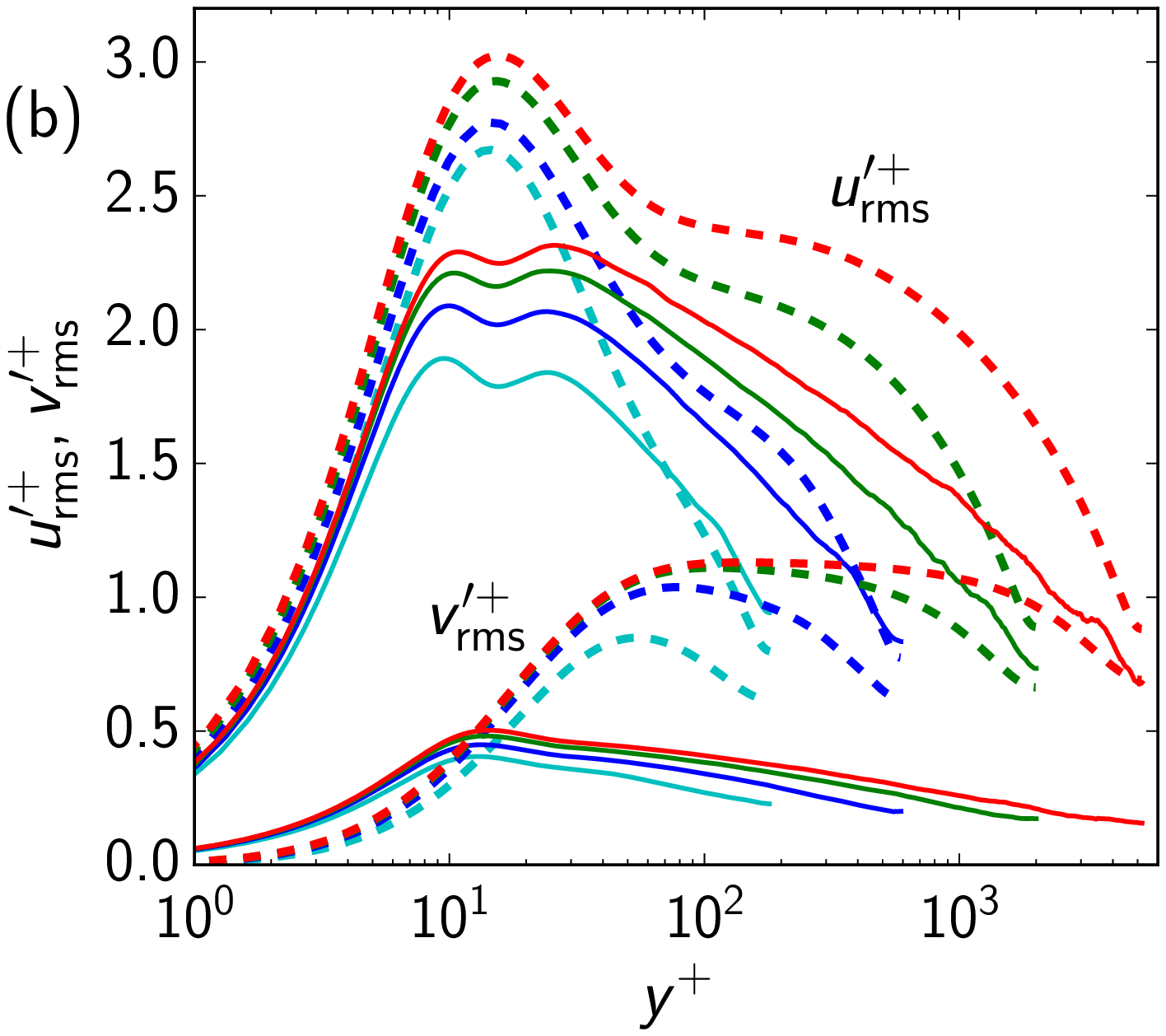}
  \caption{%
    (a) Mean streamwise velocity, $U^+$ (offset by $\Delta U^+=3$ for visibility), and
    (b) streamwise, $u^{\prime +}_\mathrm{rms}$, and wall-normal, $v^{\prime +}_\mathrm{rms}$, r.m.s. fluctuation velocity components for various $Re_\tau$ numbers for which reference data is available. 
    Reference DNS results are from \cite{Lee_Moser:2015,Moser_etal:1999}.
    Empirical profiles are described by the coefficients $\kappa=0.389$ and $B=4.23$ \cite{Marusic_etal:2010}.
  }
  \label{fig:vel}
\end{figure}

We calibrated the model by matching the velocity boundary layer of a reference DNS \cite{Lee_Moser:2015} at $Re_\tau=5200$ as described in \cite{Schmidt_etal:2003,Rakhi_etal:2019}.
This yielded the model parameters $C=6$, $Z=300$, and $\alpha=1/6$ that are kept fixed for the rest of this study.
The predictability of the model is addressed in figures~\ref{fig:vel}(a,b) in which we show the normalized mean velocity, $U^+=U/u_\tau$, and the normalized root-mean-square (r.m.s.) fluctuation velocity components, $u_{i,\mathrm{rms}}^{\prime +}=(\overline{u_i^2} - U_i^2)^{1/2} /u_\tau$, for $i=1$ ($u'$) and $i=2$ ($v'$), respectively, as function of the boundary-layer coordinate, $y^+=yu_\tau/\nu$.
The ODT mean velocity profile is in reasonable agreement with reference DNS \cite{Moser_etal:1999,Lee_Moser:2015} and the empirical law-of-the-wall \cite{Marusic_etal:2010} but lacks some features of the buffer and outer layer.
The ODT r.m.s. profiles are degraded in comparison to the reference DNS which is a known modeling artifact \cite{Schmidt_etal:2003,Lignell_etal:2013,Rakhi_etal:2019}.

\section{\label{sec:results} Results and discussion}


\subsection{\label{sec:Kp} $Sc$ and $Re_\tau$ number dependence of the scalar transfer to a wall}

\begin{figure}[t]
  \centering
  \includegraphics[height=42mm]{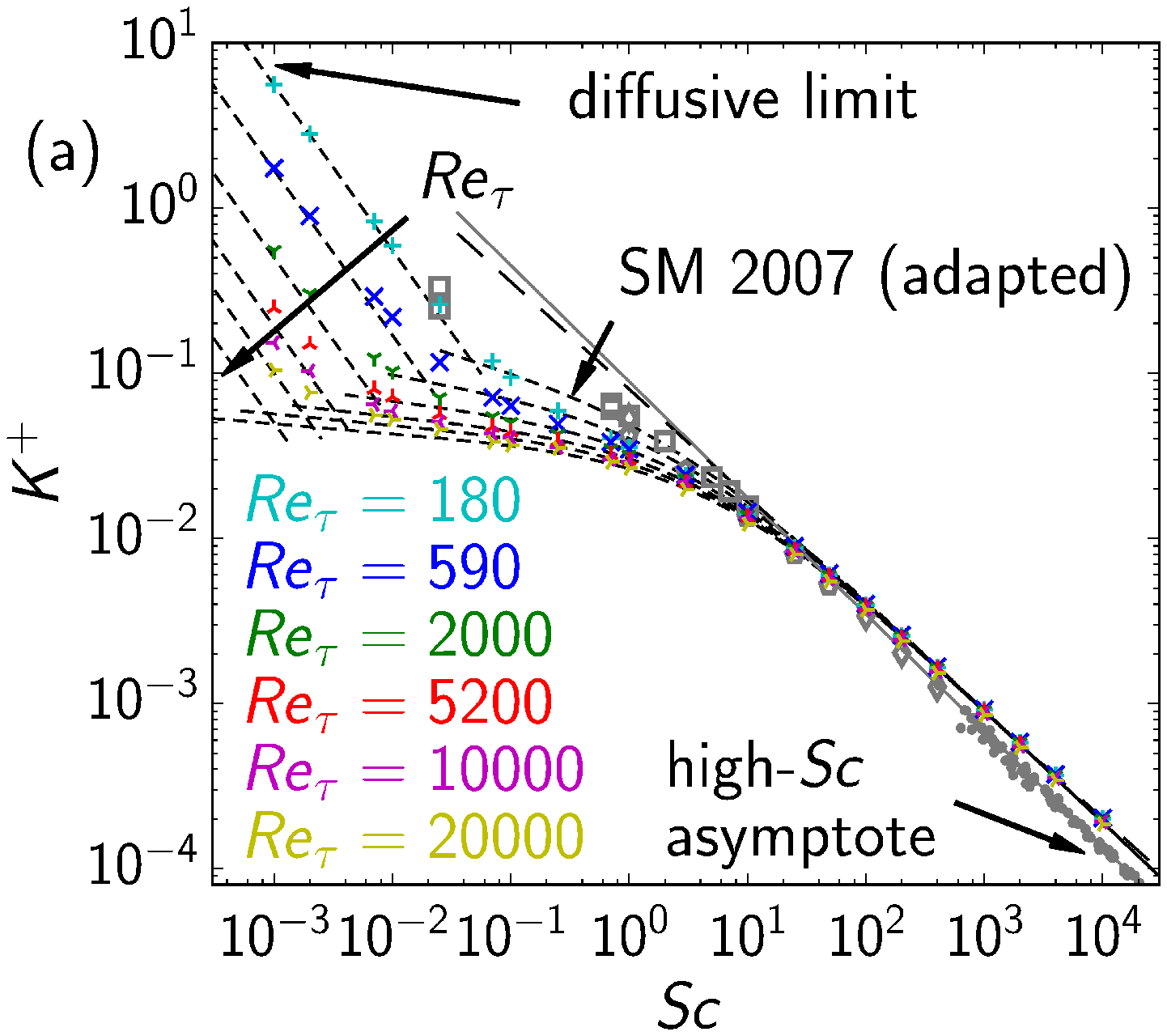} \hspace*{-3mm}
  \includegraphics[height=42mm]{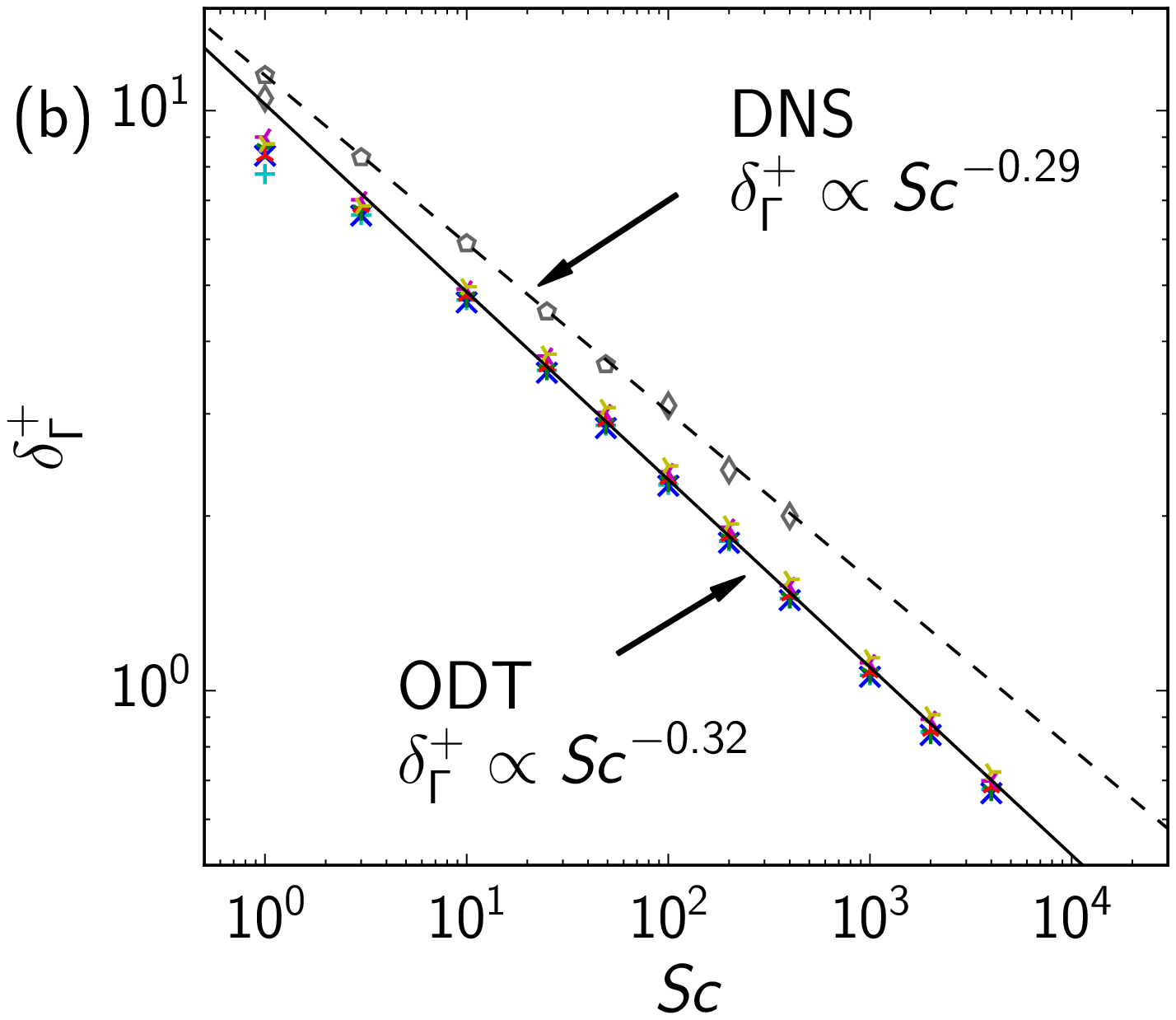} \hspace*{-3mm}
  \includegraphics[height=42mm]{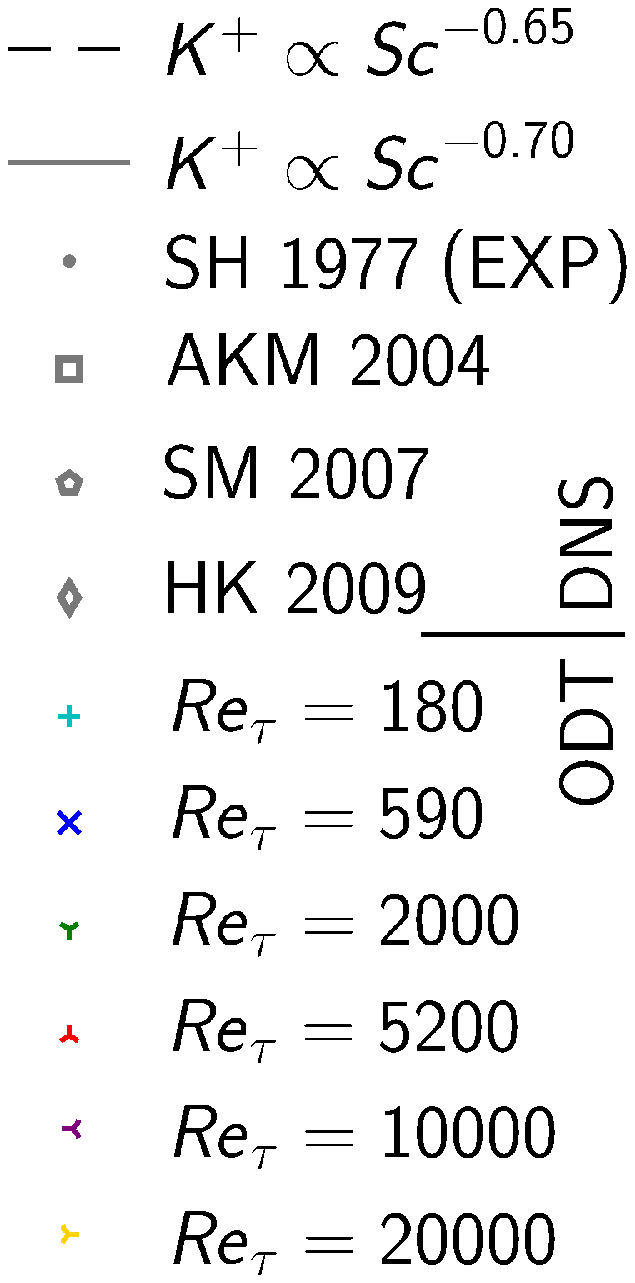}
  \caption{%
    (a) Scalar transfer coefficient, $K^+$, and 
    (b) diffusive sublayer thickness, $\delta_\Gamma^+$, for various $Sc$ and $Re_\tau$ numbers.
    The sublayer thickness $\delta_\Gamma^+$ is given by the intersection of the linear and log layer for $Sc\geq1$ (compare with figure~\ref{fig:tmean}(b) below).
    Broken and solid lines indicate empirical scaling laws.
    High $Re_\tau$ and $Sc$ number reference measurements are from \cite{Shaw_Hanratty:1977}.
    Low $Re_\tau$ and moderate $Sc$ number reference DNS are from \cite{Abe_Kawamura_Matsuo:2004,Hasegawa_Kasagi:2009,Schwertfirm_Manhart:2007}.
  }
  \label{fig:Kp}
\end{figure}

Scaling regimes of the transfer to a wall are of interest for various applications with heat or mass transport and quantified by the scalar transfer coefficient,
\refstepcounter{equation}
$$
  K^+ = Sh\big/(\gamma\,\,Sc\,Re_\tau) 
  = \theta_\tau\big/\Delta\theta \;,
  \eqno{(\theequation{\mathit{a},\mathit{b}})}
  \label{eq:Kp}
$$
where $Sh$ denotes the Sherwood (Nusselt) number, $\gamma$ a geometrical proportionality constant, $\theta_\tau=(\Gamma/u_\tau)\,\left|{\mathrm{d}\Theta}/{\mathrm{d}y}\right|_\mathrm{w}$ the friction concentration, and $\Delta\theta=|\theta_\mathrm{b}-\theta_\mathrm{w}|=|\theta_\mathrm{top}-\theta_\mathrm{bot}|/2$ the bulk-wall scalar concentration difference.
$K^+$ is shown in figure~\ref{fig:Kp}(a) and discussed in the following.

For very low $Sc$ numbers, molecular-diffusive processes dominate over turbulent advection so that $Sh\simeq1$. 
Hence, equation~(\ref{eq:Kp}\textit{a}) yields $K^+\propto Sc^{-1}Re_\tau^{-1}$ for the diffusive limit.
ODT exhibits this limit exactly, whereas reference DNS \cite{Abe_Kawamura_Matsuo:2004} for $Sc=0.025$ with $Re_\tau=395$ and $640$ yield somewhat larger values for $K^+$.
These reference DNS also seem to be less sensitive to the $Re_\tau$ number than present ODT results presumably due to details of the scalar forcing used.

For very high $Sc$ numbers, asymptotic one-dimensional theory for the mean scalar conservation equation is used with neglect of the thin asymptotic diffusive surface layer.  This yields $K^+ \propto Sc^{-(n-1)/n}$ with theoretically estimated $n=3$ or $4$ \cite{Shaw_Hanratty:1977,Son_Hanratty:1967}.
Available measurements \cite{Shaw_Hanratty:1977} for $700\leq Sc\leq37{,}000$ are consistent with an effective scaling, $K^+\propto Sc^{-0.70}$, that is, $n=3.38$.
Available reference DNS \cite{Hasegawa_Kasagi:2009} up to $Sc=400$ for much lower $Re_\tau$ number also approach this limit.
Present ODT results reach up to $Sc=4000$ and these results are well described by $K^+\propto Sc^{-0.65}$, that is, $n=2.85$ for $Sc>100$ investigated.
We will come back to this below in section~\ref{sec:edd} for the turbulent eddy viscosity.

For intermediate $Sc$ numbers, an empirical relation has been derived that accounts for overlapping linear and log layers \cite{Schwertfirm_Manhart:2007}. 
This relation is given by
\begin{equation}
  K^+(Sc,Re_\tau) = \left[
          \kappa_\theta^{-1}\ln(Re_\tau)
        + \xi\, \mathit{Sc}^{1-r}
        + r\,\kappa_\theta^{-1}\ln(Sc)
        - \kappa_\theta^{-1} \ln(\xi)
        \right]^{-1} \;,
  \label{eq:Kp-SM}
\end{equation}
where $\kappa_\theta$ is the von~K\'arm\'an constant of the scalar, and $\xi$ and $r$ are high-$Sc$-number scaling parameters of the diffusive boundary-layer thickness, $\delta_\Gamma^+\simeq \xi\, Sc^{-r}$, which is shown in figure~\ref{fig:Kp}(b).
Here, $\delta_\Gamma^+$ is obtained from the intersection of the linear and log law as shown in figure~\ref{fig:tmean}(b) below.
ODT predictions of $K^+$ shown in figure~\ref{fig:Kp}(a) are well described by the low-$Re_\tau$-number fit of equation~\eqref{eq:Kp-SM} with $\kappa_\theta=0.27$, $\xi=9.7$, and $r=0.32$ (curved dashed lines in figure~\ref{fig:Kp}(a)). 
These values are close to reference DNS \cite{Schwertfirm_Manhart:2007}, which yielded $\kappa_\theta=0.27$, $\xi=11.5$, and $r=0.29$ (not shown).
We remind that equation~(\ref{eq:Kp-SM}) is strictly valid only for $Sc\geq1$, but it is used here to provide a bound for the $Re_\tau$ number dependence of $K^+$.
ODT results for $Sc\leq O(1)$ are surprisingly well described by the empirical relation.
This suggests that the main difference between ODT and DNS is in the exponent $r$ that governs the high-$Sc$-number asymptote of $K^+$.

\subsection{\label{sec:tmean} Boundary-layer structure of the mean scalar concentration}

\begin{figure}[t]
  \centering
  \includegraphics[height=42mm]{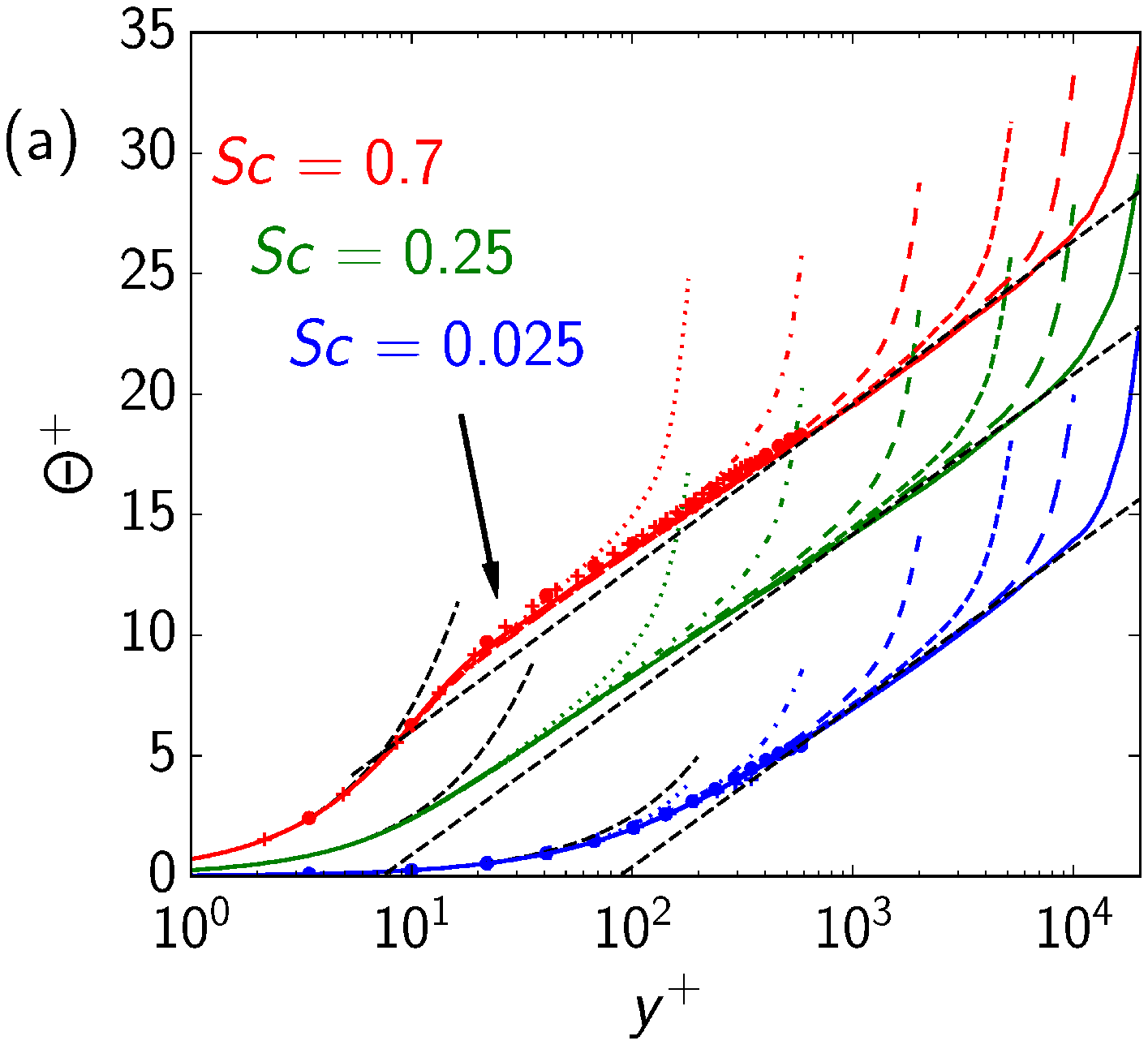} \hspace*{-3mm}
  \includegraphics[height=42mm]{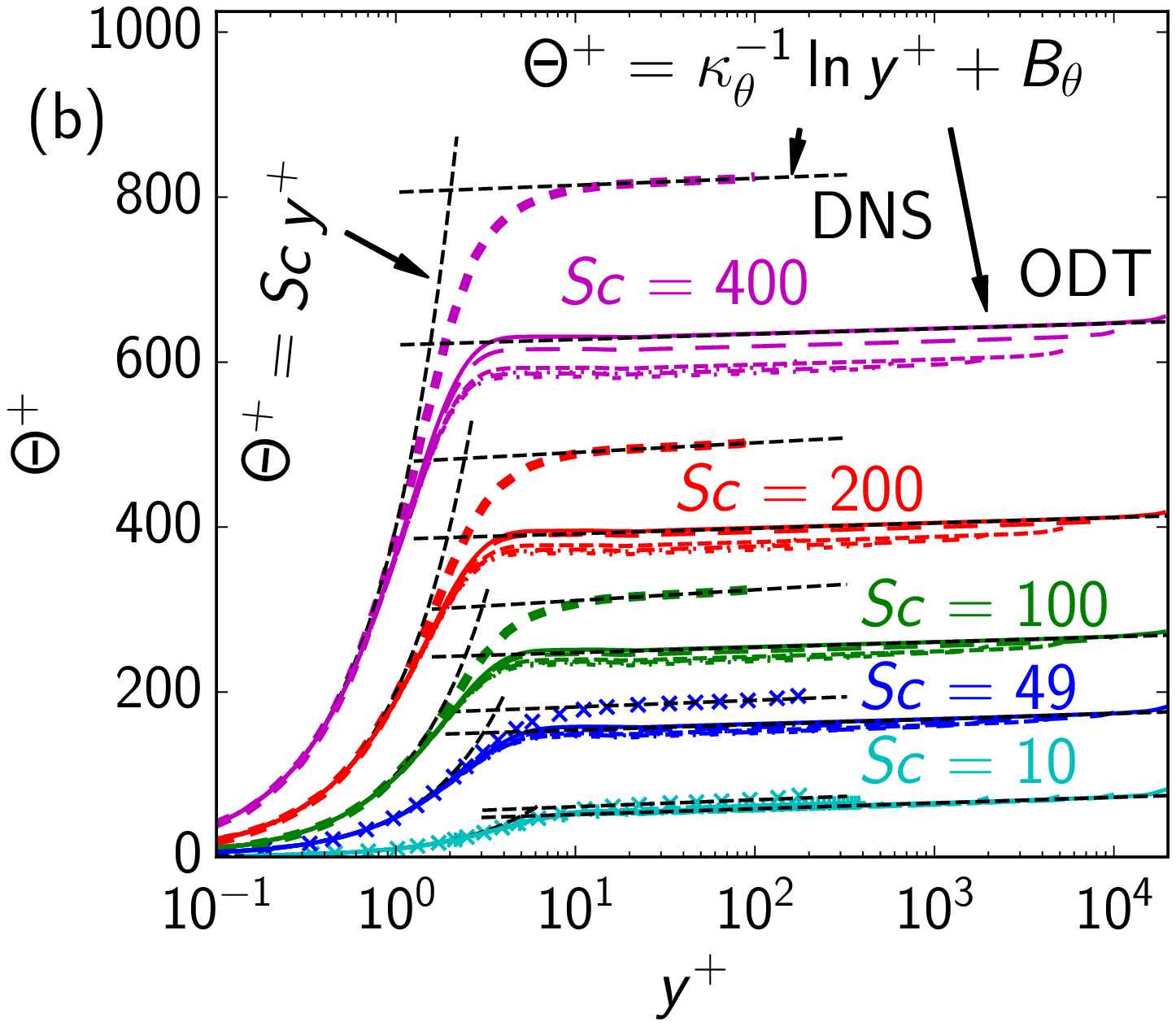} \hspace*{-3mm}
  \includegraphics[height=42mm]{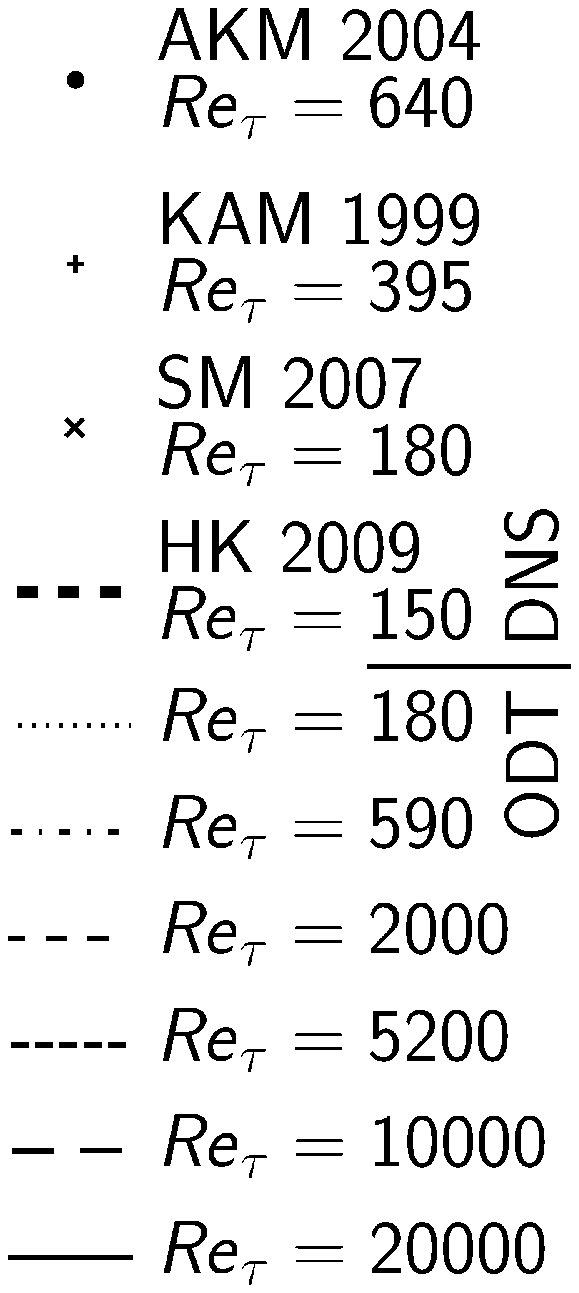}
  \caption{%
    Profiles of the normalized mean scalar concentration, $\Theta^+$, for various $Sc$ and $Re_\tau$ numbers.
    (a) Diffusion-dominated regime with $Sc<1$, and
    (b) inertia-dominated regime with $Sc>1$.
    Reference DNS results are from \cite{Abe_Kawamura_Matsuo:2004,Hasegawa_Kasagi:2009,Kawamura_Abe_Matsuo:1999,Schwertfirm_Manhart:2007}.
  }
  \label{fig:tmean}
\end{figure}

Wall-normal profiles of the mean scalar concentration, $\Theta^+=(\bar{\theta}-\theta_\mathrm{w})/\theta_\tau$, are shown in figure~\ref{fig:tmean} for various $Sc$ and $Re_\tau$ numbers.
The overlap layer exists for all $Sc\geq0.7$ investigated, but a log layer can be discerned for all cases with large friction Peclet number, $Pe_\tau=Sc\,Re_\tau\gg O(1)$. 
All log-layer fits shown have been obtained across $40\leq y^+\leq Re_\tau/2$ for the largest available $Re_\tau$ number.

For $Sc\leq0.7$ in figure~\ref{fig:tmean}(a), ODT predictions of the inner layer for $Re_\tau\geq590$ investigated are consistent with corresponding reference DNS \cite{Abe_Kawamura_Matsuo:2004,Kawamura_Abe_Matsuo:1999}.
In these DNS, the scalar is prescribed by a forcing term which affects the outer layer \cite{Pirozzoli_etal:2016}.

For $Sc\geq10$ in figure~\ref{fig:tmean}(b), the model predicts the expected boundary-layer structure but with a lower additive constant, $B_\theta$.
This is indicative of an overestimation of the wall-normal transport in the model that manifests itself by an overestimation of the mean wall gradient, $|\mathrm{d}\Theta/\mathrm{d}y|_\mathrm{w}$.
Latter is consistent with the observed overestimation of $K^+$ by virtue of equation~(\ref{eq:Kp}\textit{b}).
While $\theta_\tau$ is too large, $\Delta\theta$ is prescribed by boundary conditions and $u_\tau$ by the mean pressure gradient.
The log-layer fits shown are for the ODT predictions with $Re_\tau=20{,}000$ and these are described by the von~K\'arm\'an constant $\kappa_\theta=0.35\pm0.01$ for $Sc\geq10$ investigated.
This is closer to $\kappa\approx0.4$ of the velocity boundary layer \cite{Marusic_etal:2010} than $\kappa_\theta=0.47$ suggested by Kader \cite{Kader:1981}.
Note that a fit of the ODT results for $Re_\tau=180$ yields $\kappa_\theta=0.24\pm0.03$ which is consistent with DNS \cite{Hasegawa_Kasagi:2009,Schwertfirm_Manhart:2007}.

\subsection{\label{sec:edd} Limiting relation of the turbulent eddy diffusivity in the vicinity of a wall}

We noted in section~\ref{sec:Kp} that the thin diffusive surface layer has negligible contribution to high-$Sc$-number scalar transfer.
The $Sc$ number dependence of the scalar transfer coefficient, $K^+\propto Sc^{-(n-1)/n}$, is, hence, governed by the near-wall structure of the turbulent eddy diffusivity, $\Gamma_\mathrm{t} =  -\overline{v'\theta'} \big/ (\mathrm{d}\Theta/\mathrm{d}y)$.
This diffusivity is the ensemble effect of turbulent eddies, $-\overline{v'\theta'}$, divided by the mean scalar gradient, $\mathrm{d}\Theta/\mathrm{d}y$, and obeys inner scaling \cite{Son_Hanratty:1967}, $\Gamma_\mathrm{t}^+ = \Gamma_\mathrm{t}/\nu \propto y^{+\,n}$.

\begin{figure}[t]
  \centering
  \includegraphics[height=42mm]{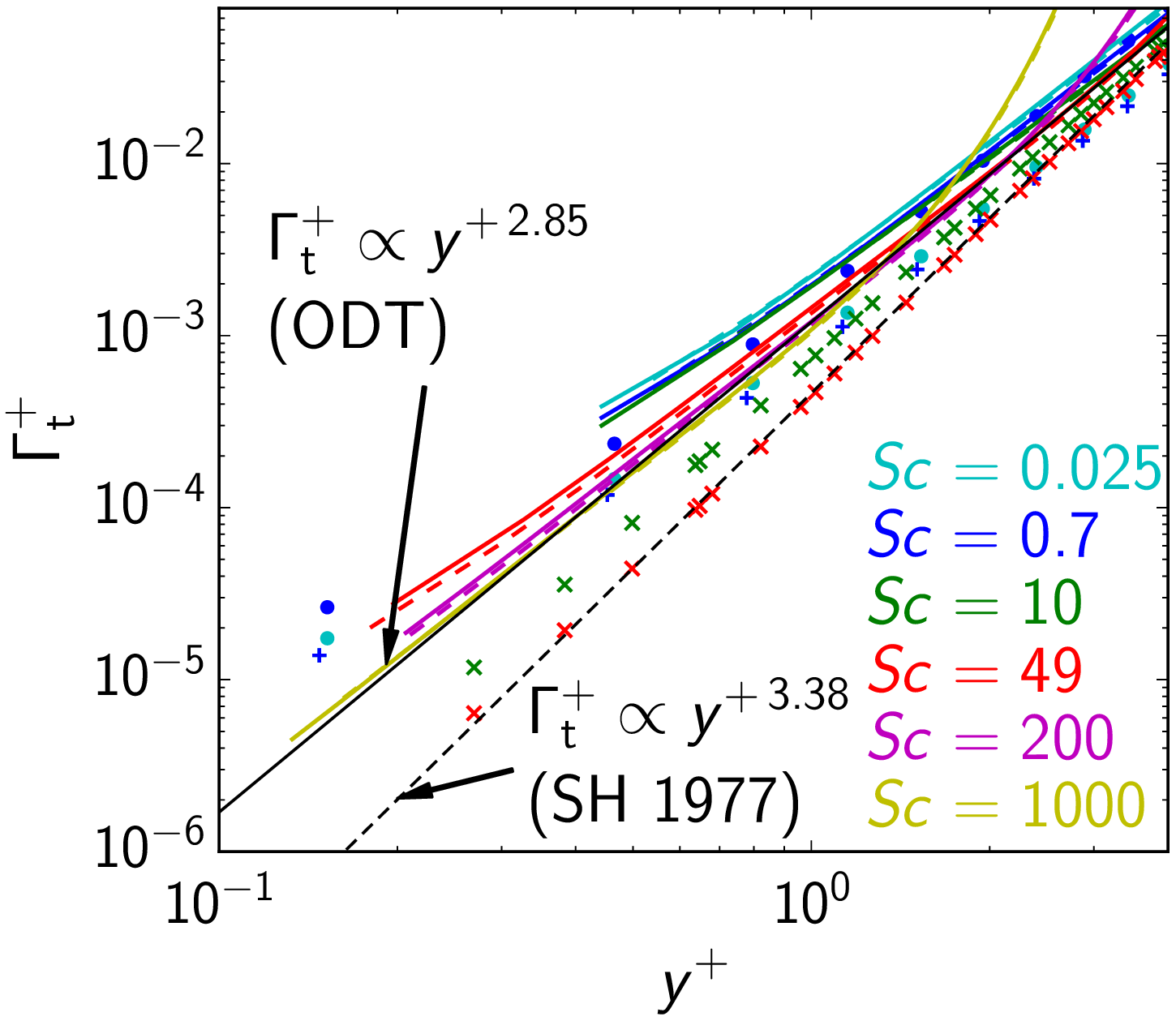} \hspace*{-3mm}
  \includegraphics[height=42mm]{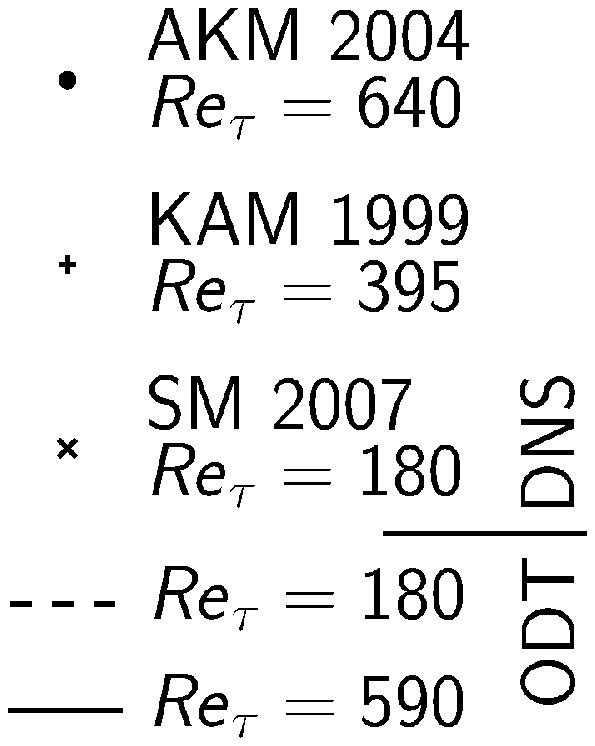}
  \caption{%
    Normalized turbulent eddy diffusivity, $\Gamma_\mathrm{t}^+=\Gamma_\mathrm{t}/\nu$, in the vicinity of the wall for various $Sc$ and $Re_\tau$ numbers and its limiting relation.
    Only data for $Re_\tau\leq640$ investigated are shown to aid visibility.
    Reference DNS results are from \cite{Abe_Kawamura_Matsuo:2004,Kawamura_Abe_Matsuo:1999,Schwertfirm_Manhart:2007}.
  }
  \label{fig:edd}
\end{figure}

Figure~\ref{fig:edd} shows $\Gamma_\mathrm{t}^+$ in the vicinity of the wall together with high-$Sc$-number limiting relations.
Reference data \cite{Hasegawa_Kasagi:2009,Schwertfirm_Manhart:2007,Shaw_Hanratty:1977} for high $Sc$ numbers is consistent with $n=3.38$, that is, $K^+\propto Sc^{-0.70}$ shown in figure~\ref{fig:Kp}(a).
Present ODT results exhibit $n=2.85\pm0.05$, which is consistent with $K^+\propto Sc^{-0.65}$ and close to asymptotic one-dimensional theory with $n=3$ \cite{Shaw_Hanratty:1977}.
For $Sc\approx50$, present ODT results more closely resemble available reference DNS \cite{Abe_Kawamura_Matsuo:2004,Kawamura_Abe_Matsuo:1999} for $Sc\leq0.7$. 

\section{\label{sec:conc} Conclusions}

Numerical simulations of passive scalars in turbulent channel flows have been performed for a relevant range of Schmidt and Reynolds numbers utilizing an adaptive implementation of the one-dimensional turbulence model.
The model simultaneously resolves all fluctuating flow variables on all relevant scales, captures the turbulent boundary layer, and exhibits consistent low-order flow statistics.
The predicted scalar transfer to a wall is consistent with asymptotic one-dimensional theory, which implies a weak but systematic overestimation of the high-Schmidt-number asymptote. 
Altogether, the model has good predictive capabilities, which is an important property for forward modeling of heat, mass, and momentum transfer in multi-physics turbulent boundary layers.

\section*{Acknowledgements}
 
We thank Alan Kerstein for commenting on an early version of this manuscript.

Funding: This work was supported by the European Regional Development Fund [grant number StaF~23035000].





\end{document}